\documentclass[twocolumn,preprintnumbers,amsmath,amssymb,prl,showpacs,superscriptaddress]{revtex4}

\usepackage{graphicx}
\usepackage{dcolumn}
\usepackage{bm}

\begin{document}
\title{Dynamical Heterogeneity and Nonlinear Susceptibility in Short-Ranged Attractive Supercooled Liquids}
\author{P. Charbonneau}\affiliation{Chemistry and Chemical Biology, Harvard
University, 12 Oxford Street, Cambridge, Massachusetts 02138,
USA~\footnote{Current address: FOM Institute for Atomic and
Molecular Physics, Kruislaan 407, 1098 SJ Amsterdam, The
Netherlands.}}
\author{D. R. Reichman}\email{drr2103@columbia.edu} \affiliation{Department of
Chemistry, Columbia University, 3000 Broadway, New York, New
York 10027, USA} 
\begin{abstract}
Recent work has demonstrated the strong qualitative differences
between the dynamics near a glass transition driven by
short-ranged repulsion and one governed by short-ranged
attraction.  Here, we study in detail the behavior of
non-linear, higher-order correlation functions that measure the
growth of length scales associated with dynamical heterogeneity
in both types of systems.  We find that this measure is
qualitatively different in the repulsive and attractive cases
with regards to the wave vector dependence as well as the time
dependence of the standard non-linear four-point dynamical
susceptibility.  We discuss the implications of these results
for the general understanding of dynamical heterogeneity in
glass-forming liquids.
\end{abstract} \maketitle

The underlying reasons for the dramatic increase in the
viscosity of glass-forming liquids are not well understood.  It
has become increasingly clear that simple structural measures
remain short-ranged close to the glass transition, and thus a
growing simple static length scale does not appear to be
implicated~\cite{ediger:1996}. This has led to the search for a
growing dynamical length scale that drives vitrification.
Indeed, recent
simulations~\cite{yamamoto:1998,Donati:1998,donati:1999,toninelli:2005,widmer:2005,widmer:2004}
and
experiments~\cite{cicerone:1995,bohmer:1998,Russell:2000,weeks:2000,richert:2002,keys:2007}
have given direct evidence for both a growing length scale and
a dynamical scaling relating its growth to the rapidly
increasing time scales that characterize the glass transition.
The study of this key aspect of dynamical heterogeneity, as
encoded in various multi-point dynamical susceptibilities, has
opened up the ability both to extract absolute length scales
associated with cooperative relaxation in glassy systems and to
provide precise metrics for the testing of various theoretical
approaches~\cite{berthier:2005,chandler:2006,toninelli:2005}.

The simplest model system that exhibits the expected dynamical
behavior associated with more complicated glassy systems is the
hard-sphere liquid.  Here, entropy-driven crowding effects give
rise to a characteristic dynamical behavior that includes a
two-step non-exponential relaxation, a dramatic increase in
relaxation times associated with small changes in volume
fraction, and dynamical heterogeneity accompanied by a growing
dynamical length scale~\cite{berthier:2005}.  Recently, it has
been demonstrated via theory~\cite{dawson:2001},
simulation~\cite{puertas:2003,zaccarelli:2002}, and
experiment~\cite{chen:2003,pham:2002} that another extreme
glassy limit exists for simple spherical particles: that of the
short-range attractive glassy state. Here, strong short-ranged
bonding between the particles can lead to extremely slow
relaxation, but with dramatically different dynamical
characteristics.  Dynamical heterogeneity also exists close to
the attractive glassy
state~\cite{puertas:2003,puertas:2004,puertas:2005,reichman:2005},
but has not been systematically characterized, and multi-point
dynamical susceptibilities for such systems have not been
measured or computed.  The goal of this work is to investigate
in detail the properties of standard non-linear spatiotemporal
susceptibilities at distinct points along the attractive glass
line and to quantitatively and qualitatively compare the
observed features to those of the hard-sphere system that lies
at the infinite temperature limit of the dynamical arrest line.
The outcome of this exercise is a greater understanding of the
physics of attractive glass-forming systems and dynamical
heterogeneity in glass-forming liquids in general. Further,
this work provides important benchmarks for the testing of
various theoretical approaches.

\begin{figure}
\center{\includegraphics[width=\columnwidth]{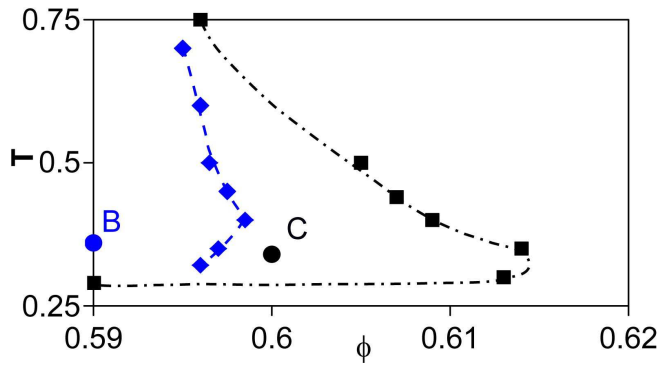}}
\caption[Dynamical phase diagram]{(Color online) Dynamical
phase diagram for interaction ranges $n=30$ (diamonds) and
$n=40$ (squares) obtained by extrapolating to the limit of zero
diffusivity~\cite{sciortino:2003}. The lines are guides for the
eye. The putative higher-order ($A_4$) singularity lies near
the $n=30$ cusp. Simulation points $B$ and $C$ are indicated,
while point $A$ lies off the plot in the high-$T$ limit of the
$T$-$\phi$ diagram. Dashed lines are guides for the eye.}
\label{fig:dpd}
\end{figure}

The systems that we consider have potentials of the form
\begin{equation}
U(r)=4\varepsilon
\left[\left(\frac{\sigma_{\alpha\beta}}{r}\right)^{2n}-
\left(\frac{\sigma_{\alpha\beta}}{r}\right)^n\right],
\end{equation}
where the temperature scale $T$ is set by $\varepsilon$, the
length scale is set by $\sigma_{BB}$ and time $t$ is rescaled
by $(\varepsilon/m\sigma_{AA}^2)^{1/2}$. For our study $n=40$
and $30$, yielding a potential with an attractive range of
approximately $3\%$ and $4\%$ of $\sigma_{BB}$, respectively.
To prevent crystallization, a 50:50 mixture with size ratio
$\sigma_{AA}/\sigma_{BB}=1.2$ and
$\sigma_{AB}=\sigma_{BA}=(\sigma_{AA}+\sigma_{BB})/2$ is used.
Standard molecular dynamics simulations with a number of
particles $N=256$ have been performed in the microcanonical
ensemble with a time step $\Delta t<1.3\times 10^{-3}$.
Finite-size effects were tested by comparison to a $N=2048$
system with little discernable difference found for the
quantities studied here. In Fig.~\ref{fig:dpd} we plot a
dynamical phase diagram ($T$ vs. volume fraction $\phi$) of the
system. The arrest line is determined by extrapolation of the
iso-diffusion curves to the limit of zero
diffusivity~\cite{sciortino:2003}. Three ($T,\phi$,$n$) state
points in the supercooled-liquid regime are considered:
$A=(4.4,0.605,30)$, $B=(0.36,0.59,30)$, and $C=(0.34,0.6,40)$.
Point $A$ lies close to the hard-sphere limit, while point $C$
lies close to the attraction-driven arrest line, but away from
the putative ($A_3$) dynamical singularity predicted by
mode-coupling theory (MCT)~\cite{gotze:2002}. Point $B$ lies
close to the arrest line in the ``reentrant pocket'', near the
location of the higher-order ($A_4$) singularity predicted by
MCT.  In all cases, state points have been chosen not only to
reflect potentially distinct physics, but so that the
$\alpha$-relaxation time $\tau_{\alpha}$ and the bulk diffusion
constants are similar as well~\cite{footnote1}.
This allows for comparison of the potentially
different physics at points $A$-$C$ for comparable degrees of
absolute sluggishness.

\begin{figure}
\center{\includegraphics[width=\columnwidth]{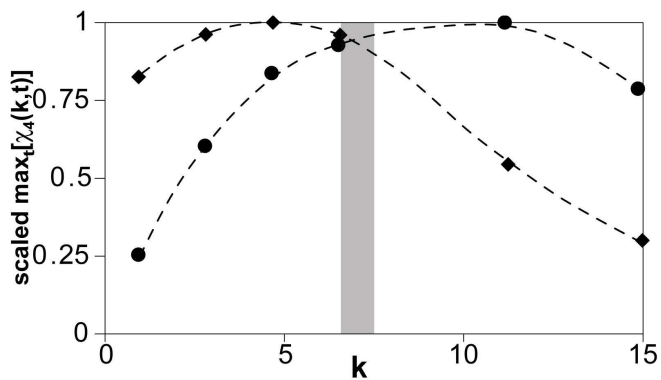}}
\caption[Maximal values of $\chi_4(k,t)$ for different
regimes]{Maximal values of $\chi_4(k,t)$ as a function of wave
vector for points $A$ (diamonds) and $C$ (circles), relative to
the absolute maximum of  $\chi_4(k,t)$ for that state point.
The $k$-dependence of point $B$ is qualitatively similar to
that of point $C$ and is not shown. The nearest-neighbor peak
of $S(k)$ for both state points is contained within the shaded
area. Dashed lines are guides for the eye.} \label{fig:maxchi4}
\end{figure}

Simple two-point dynamical correlation functions have already
been extensively characterized in these
systems~\cite{sciortino:2003}. The two-point function
$F_s(k,t)$ displays drastically different decay characteristics
at the three selected state points, ranging from standard
two-step relaxation with associated power-law relaxation at
point $A$ to intermediate-time logarithmic decay spanning
several time decades at point $B$. Here we characterize the
{\em fluctuations} of two-point dynamical quantities that are
directly relevant for understanding dynamical heterogeneity. In
particular, we focus on the normalized $\chi_4(k,t)$
susceptibility defined as
\begin{equation}
\chi_{4}(k,t)=\frac{\langle f_s(k,t)^2\rangle-\langle f_s(k,t)
\rangle^2}{N^{-1}\sum_i\left[\langle
f_s^i(k,t)^2\rangle-\langle f_s^i(k,t)\rangle^2\right]},
\end{equation}
where $f_s(q,t)=N^{-1}\sum_i
\cos{\left\{\mathbf{k}\cdot\left[\mathbf{r}_i(t)-\mathbf{r}_i(0)\right]\right\}}\equiv
N^{-1}\sum_i f_s^i(k,t)$.  This quantity measures the size of
fluctuations in self-density correlations at a particular wave
vector; its growth is related to the increase of a cooperative
dynamical length scale.  Here, unlike in some earlier work, we
explicitly label the wave vector dependence of this
susceptibility, which indicates that the fluctuations
associated with dynamical heterogeneity may be large or small
depending on the {\em intrinsic} length scale that is probed.
Indeed, as discussed by Chandler {\em et
al.}~\cite{chandler:2006} (see also Ref.~\cite{dauchot:2005}),
the $k$-dependence of $\chi_4(k,t)$ is a useful way to probe
the various length scales associated with dynamical
heterogeneity. The $k$-dependence of $\chi_4(k,t)$ should not
be confused with that of $S_4(k,t)$, which provides a
four-point analog to the static structure factor, and allows
for the {\em direct} extraction of a length scale associated
with cooperative heterogeneous motion.  On the other hand, it
is expected that $max_t\left[\chi_4(k,t)\right]\sim
\xi(t)^{2-\eta}$, where $\xi(t)$ is the same dynamical
heterogeneity length scale extracted from
$S_4(k,t)$~\cite{lacevic:2003} and $\eta$ is the susceptibility
exponent. Thus, as long as $\eta$ does not vary for the region
of interest in the dynamical phase diagram, one may infer some
information concerning the growth and absolute size of
$\xi(t)$~\cite{footnote2}.
Lastly, it should be noted that the definition of $\chi_4(k,t)$
given above slightly differs from the standard definition, due
to the normalization factor in the denominator. This
normalization is used to attempt an unbiased comparison of peak
amplitudes. We have checked that the conclusions drawn from the
results presented below are not altered if the standard,
unnormalized definition is used.

We start with a comparison between the $k$-dependence of the
maximal values of $\chi_4(k,t)$ for state points $A$ and $C$.
As shown in Fig.~\ref{fig:maxchi4}, a striking qualitative
distinction exists between the size of dynamical fluctuations
in the cases where glassy behavior is driven by strong,
short-ranged bonding compared to the hard-sphere limit, where
crowding drives vitrification.  In particular, dynamical
fluctuations are maximized for wave vectors {\em below} that of
the main diffraction peak of $S(k)$ in the hard-sphere limit,
while in the attraction-driven case the maximal fluctuations
occur for wave vectors {\em in excess} of the first-neighbor
peak of $S(k)$. This finding makes clear the fact that, while
in the hard-sphere case dynamical heterogeneity fluctuations
are most sensitive to collective events on scales larger than
the particle size, in the case of strong short-ranged
attractions it is bonding fluctuations that trigger the
emergence of dynamical heterogeneity (potentially associated
with large length scales), as measured in $\chi_4(k,t)$.

It is instructive to compare this result with the recent
calculations of Greenall~{\em et al.}~\cite{greenall:2006}. In
this work, the sensitivity of the $k$-dependent plateau height
to changes in the structure, as computed by MCT, was measured
for both the hard-sphere and the attractive glass-forming
limits. For the hard-sphere system, it was found that
sensitivity is most pronounced for changes in structure {\em
just beyond} the first shell of neighbors.  Greenall~{\em et
al.} deemed this the ``caged-cage'' effect.  On the other hand,
it was found that the plateau for systems near the
attraction-driven arrest line is most sensitive to changes of
structure that occur at high wave vectors associated with
short-ranged bonding -- i.e. at $k$ values much in excess of an
inverse particle size.

A strong qualitative similarity is thus seen between the
$k$-dependence of the plateau sensitivity, as computed by MCT,
and the $k$-dependence in the peak height of $\chi_4(k,t)$ as
measured directly via MD simulation.  To interpret this we
first remark that, at least in the $\beta$-relaxation regime,
$\chi_4(k,t)$ is an indicator of the spatial fluctuations of
the plateau height, which has been shown to be spatially
heterogeneous even at these short time scales in recent
numerical simulations~\cite{widmer:2006}. Clearly the MCT
calculations of Greenall~{\em et al.} measure the sensitivity
of the plateau of a {\em uniform} system to {\em uniform}
changes in structure, while at any instant in a real liquid the
local structure varies from site to site. However, it is
reasonable to assume that these {\em spatial} fluctuations will
mirror the very same sensitivity to local structure as the
global plateau does to a global change in structure.  Thus, our
dynamical results for the $k$-dependence of the peak height of
fluctuations associated with dynamical heterogeneity provides a
deep connection with the static MCT calculations of Greenall
{\em et al.} This interpretation is in harmony with recent
calculations and speculations concerning the nature and
interpretation of dynamical heterogeneity within
MCT~\cite{biroli:2006,biroli:2007}.

\begin{figure}
\center{\hspace{-0.02\columnwidth}\includegraphics[width=\columnwidth]{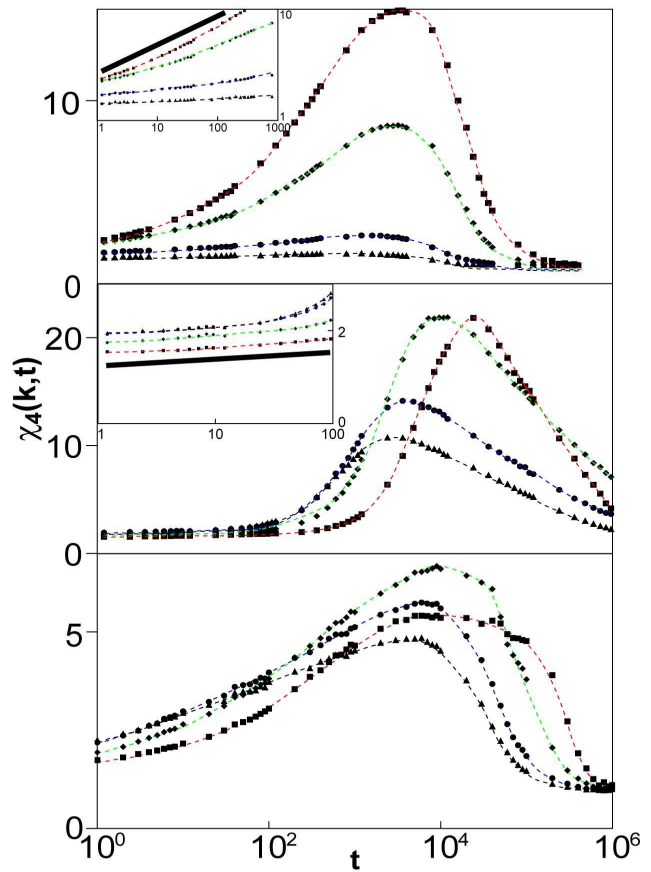}}
\caption[Evolution of $\chi_4(k,t)$ for different
regimes]{(Color online) Time evolution of $\chi_4(k,t)$ for
wave vector $k=6.6$ (squares), $11.2$ (diamonds), $18.7$
(circles), and $22.5$ (triangles) at state points $A$, $B$, and
$C$, from top to bottom. Insets: thick black lines show
approximate power-law growth at intermediate times in
hard-sphere limit (top) and logarithmic growth at state point
$B$ (middle). Dashed lines are guides for the eye.}
\label{fig:chi4}
\end{figure}

We now turn to the full time dependence of $\chi_{4}(k,t)$. In
Fig.~\ref{fig:chi4} we show $\chi_{4}(k,t)$ for $k$-values
above and below the peak of $S(k)$ for the three different
state points.  Clearly the temporal shape of $\chi_{4}(k,t)$ is
{\em qualitatively} different at the three points.  The
time-dependent growth of $\chi_{4}(k,t)$ to its peak in the
repulsion-driven limit of point $A$ may be fit to a power-law
form, as shown in the log-log inset of
Fig.~\ref{fig:chi4}(top). This behavior is similar to that
observed in many other systems, such as mixtures of soft-sphere
or Lennard-Jones particles~\cite{toninelli:2005}. This is in
contrast to the case of point $B$, which lies closest to a
putative MCT higher-order singularity.  Here, as shown in
Fig.~\ref{fig:chi4}(middle) the amplitude of $\chi_{4}(k,t)$ in
the $\beta$ regime is extremely small, while the peak heights
in the $\alpha$ regime are sizable and actually exceed those
calculated in the hard-sphere case. Thus, the $\beta$ regime at
this state point is {\em local} in its physics, displaying none
of the hallmarks of cooperativity that have already set in at
short times in the hard-sphere case. Further, the dramatic
change in temporal behavior of $\chi_{4}(k,t)$ from
intermediate to long times suggest the possibility of different
length scales governing the $\beta$ and $\alpha$ regimes,
respectively, in contrast to what is usually observed in
typical glassy systems~\cite{berthier:2007,berthier:2007b}.
Lastly, at state point $C$, where attractions dominate
relaxation but far from the location of the reentrant elbow of
the arrest line, a behavior with mixed properties is seen. In
particular, the growth of $\chi_{4}(k,t)$ does not display the
drastic difference between the $\beta$ and $\alpha$ regimes,
although the growth is quite slow, and peak values reach only
modest amplitudes.

A close inspection of the behavior of $\chi_{4}(k,t)$ at state
point $B$ indicates that it grows in the $\beta$ regime not as
a power law, but essentially logarithmically in time (see
linear-log inset of Fig.~\ref{fig:chi4}(middle)). While the
decay of the two-point function $F_s(k,t)$ is known to be
nearly logarithmic in this regime~\cite{sciortino:2003}, it is
not at all obvious that this should also be true for its {\em
fluctuations}. It has recently be argued that the
susceptibility $\chi_T(k,t)\sim\frac{dF_s(k,t)}{dT}$ may serve
as a mimic of $\chi_4(k,t)$~\cite{berthier:2005,
berthier:2007,berthier:2007b}. By considering $F_s(k,t)$, it is
clear that the growth of $\chi_4(k,t)$ to its peak should be a
power law for standard repulsive systems. On the other hand,
the same considerations are not informative for the attractive
regime, where the leading term of $\chi_T(k,t)$ in the $\beta$
regime is not dependent on time at all. Indeed, argumentation
based on the susceptibility $\chi_T(k,t)$ would suggest that
$\chi_4(k,t)$ grows logarithmically in time only if {\em
subleading} terms in the expansion of the two-point function is
a power series in the logarithm, as suggested from MCT analysis
near higher-order singularities~\cite{gotze:2002}. Indeed, the
result presented here for $\chi_4(k,t)$ may be taken as
indirect evidence for the reality of these subleading terms.

In conclusion we have systematically studied how the dynamical
heterogeneity indicator $\chi_4(k,t)$ varies along the arrest
line in attractive colloidal systems.  The behavior of
$\chi_4(k,t)$ in the attraction-dominated limit markedly
differs from that previously observed in standard
repulsion-dominated systems. First, the scale of fluctuations
is maximized for intrinsic length scales significantly smaller
than a particle diameter, as opposed to the hard-sphere case
where fluctuations are maximized at length scales in excess of
the particle size. This result suggests that short-ranged
bonding fluctuations trigger dynamical heterogeneity in
attractive systems, while intrinsic dynamics at scales larger
than the cage scale couple most strongly to dynamical
heterogeneity in repulsive systems. In addition, the time
dependence of $\chi_4(k,t)$ varies dramatically from one limit
to the other. These suggest marked differences in the degree of
cooperativity seen in attractive and repulsive cases.  In
particular, the amplitude of $\chi_4(k,t)$ is much smaller in
the $\beta$ regime in the attractive case, and the growth of
$\chi_4(k,t)$ is logarithmically slow near the onset of
reentrance, as opposed to the more common power-law behavior.
These results deepen our understanding of the physics of
dynamical heterogeneity as well as provide testable targets for
theoretical approaches. In particular, it would be interesting
to apply the recent extension of MCT of
Ref.~\cite{biroli:2006}, which has successfully predicted the
behavior of $\chi_4(k,t)$ and the growth of the dynamical
length scale in standard systems to the case of attractive
glass-forming systems.

\vspace{0.05cm} This work was supported in part by grants
No.~NSF-0134969 and No.~FQRNT-91389 (to PC). We would like to
thank G.~Biroli, M.~E.~Cates, J.~D.~Eaves, and K.~Miyazaki for
helpful discussions.
\bibliography{Chi4}

\begin{thebibliography}{35}
\bibitem[{foo({\natexlab{a}})}]{footnote1}
\bibinfo{note}{The relaxation time $\tau_{\alpha}$ is defined as the time at
  which the self-intermediate scattering function $F_s(k,t)=\left\langle
  N^{-1}\sum_i e^{i \mathbf{k}\cdot\left[\mathbf{r}_i(t)
  -\mathbf{r}_i(0)\right]}\right\rangle$ has decayed to $e^{-1}$ at the peak of
  the structure factor $S(k)=\langle N^{-1}\sum_{ij}
  e^{i\mathbf{k}\cdot\left(\mathbf{r}_j- \mathbf{r}_i\right)}\rangle$, where
  $\mathbf{r}_i$ is the position of particle $i$ and $k$ is the magnitude of
  the wave vector probed.}
\bibitem[{foo({\natexlab{b}})}]{footnote2}
\bibinfo{note}{This is only the case if $\chi_{4}(k,t)$ is computed in the
  thermodynamic ensemble consistent with experimental
  conditions~\cite{berthier:2005,berthier:2007,berthier:2007b}.}
\end{thebibliography}
\end{document}